\documentclass[]{spie}  

\usepackage{amsmath,amsfonts,amssymb,mathrsfs}
\usepackage{graphicx}
\usepackage[colorlinks=true, allcolors=blue]{hyperref}
\usepackage{astro_bib_macro} 

\usepackage{float} 

\title{Integrated photonic-based coronagraphic systems for future space telescopes}

\author[a]{Niyati Desai}
\author[b]{Lorenzo K\"onig}
\author[c]{Emiel Por}
\author[d,e]{Roser Juanola-Parramon}
\author[f]{Ruslan Belikov}
\author[g]{Iva Laginja}
\author[h,i,j,k]{Olivier Guyon}
\author[c]{Laurent Pueyo}
\author[f]{Kevin Fogarty}
\author[b]{Olivier Absil}
\author[l]{Lisa Altinier}
\author[g]{Pierre Baudoz}
\author[m]{Alexis Bidot}
\author[n]{Markus Johannes Bonse}
\author[o,p]{Kimberly Bott}
\author[q]{Bernhard Brandl}
\author[m]{Alexis Carlotti}
\author[r]{Sarah L. Casewell}
\author[l]{Elodie Choquet}
\author[s]{Nicolas B. Cowan}
\author[q,t]{David Doelman}
\author[u]{J. Fowler}
\author[n]{Timothy D. Gebhard}
\author[g,v,w]{Yann Gutierrez}
\author[h]{Sebastiaan Y. Haffert}
\author[v]{Olivier Herscovici-Schiller}
\author[m]{Adrien Hours}
\author[q]{Matthew Kenworthy}
\author[q]{Elina Kleisioti}
\author[x]{Mariya Krasteva}
\author[q]{Rico Landman}
\author[m]{Lucie Leboulleux}
\author[g]{Johan Mazoyer}
\author[y]{Maxwell A. Millar-Blanchaer}
\author[m]{David Mouillet}
\author[z]{Mamadou N’Diaye}
\author[q]{Frans Snik}
\author[q]{Dirk van Dam}
\author[h]{Kyle van Gorkom}
\author[$\alpha$]{Maaike van Kooten}
\author[$\beta$]{Sophia R. Vaughan}

\affil[a]{Department of Astronomy, California Institute of Technology, Pasadena, CA, USA}
\affil[b]{STAR Institute, Universit\'e de Li\`ege, Li\`ege, Belgium}
\affil[c]{Space Telescope Science Institute, Baltimore, MD, USA}
\affil[d]{NASA Goddard Space Flight Center, Greenbelt, MD, USA}
\affil[e]{University of Maryland Baltimore County, 1000 Hilltop Cir, Baltimore, MD, USA}
\affil[f]{NASA Ames Research Center, Moffett Field, USA}
\affil[g]{LESIA, Observatoire de Paris, Universit\'e PSL, CNRS, Sorbonne Universit\'e, Universit\'e de Paris, Meudon, France}
\affil[h]{Steward Observatory, University of Arizona, Tucson, AZ, USA}
\affil[i]{Subaru Telescope, NAOJ, USA}
\affil[j]{College of Optical Sciences, University of Arizona, Tucson, AZ, USA}
\affil[k]{Astrobiology Center, Osawa, Mitaka, Tokyo, Japan}
\affil[l]{Aix Marseille Univ, CNRS, CNES, LAM, Marseille, France}
\affil[m]{Univ. Grenoble Alpes, CNRS, IPAG, Grenoble, France}
\affil[n]{Max Planck Institute for Intelligent Systems, Tübingen, Germany \& ETH Zurich, Switzerland}
\affil[o]{University of California, Riverside}
\affil[p]{NASA Nexus for Exoplanet System Science, Virtual Planetary Lab Team, Seattle, USA}
\affil[q]{Leiden Observatory, Leiden University, Leiden, The Netherlands}
\affil[r]{School of Physics and Astronomy, University of Leicester, Leicester, UK}
\affil[s]{Department of Earth \& Planetary Sciences and Department of Physics, McGill University, Montr\'eal, QC, Canada}
\affil[t]{SRON Netherlands Institute for Space Research, Leiden, The Netherlands}
\affil[u]{University of California, Santa Cruz}
\affil[v]{DTIS, ONERA, Universit\'e Paris Saclay, Palaiseau, France}
\affil[w]{DOTA, ONERA, Ch\^atillon, France}
\affil[x]{European Space Agency, ESTEC, The Netherlands}
\affil[y]{Department of Physics, University of California, Santa Barbara, CA, USA}
\affil[z]{Universit\'e C\^ote d’Azur, Observatoire de la C\^ote d’Azur, CNRS, Laboratoire Lagrange, France}
\affil[$\alpha$]{National Research Council Canada, Herzberg Astronomy and Astrophysics Research Center, Victoria, Canada}
\affil[$\beta$]{University of Oxford, UK}

\begin{document} 
\maketitle

\begin{abstract}

The detection and characterization of Earth-like exoplanets around Sun-like stars is a primary science motivation for the Habitable Worlds Observatory. However, the current best technology is not yet advanced enough to reach the $10^{-10}$ contrasts at close angular separations and at the same time remain insensitive to low-order aberrations, as would be required to achieve high-contrast imaging of exo-Earths. Photonic technologies could fill this gap, potentially doubling exo-Earth yield. We review current work on photonic coronagraphs and investigate the potential of hybridized designs which combine both classical coronagraph designs and photonic technologies into a single optical system. We present two possible systems. 
First, a hybrid solution which splits the field of view 
spatially such that the photonics handle light within the inner working angle and a conventional coronagraph that suppresses starlight outside it. Second, a hybrid solution where the conventional coronagraph and photonics operate in series, complementing each other and thereby loosening requirements on each subsystem. As photonic technologies continue to advance, a hybrid or fully photonic coronagraph holds great potential for future exoplanet imaging from space.

\end{abstract}

\keywords{High-contrast imaging, instrumentation, exoplanets, coronagraph, photonics}

\section{INTRODUCTION}

\subsection{Motivation}

NASA is embarking on an ambitious program to develop and implement the Habitable Worlds Observatory (HWO) flagship mission over the next two decades. In addition to greatly advancing general astrophysics, its driving science goal is to directly image ~25 potentially Earth-like planets and spectroscopically characterize them for signs of life, as recommended by the Astro2020 decadal survey\cite{Astro20}. However, current coronagraphic instrument designs for such a mission are inefficient compared to what is theoretically possible. This “efficiency gap” has been recently recognized by the NASA Exoplanet Exploration program “technology gap list”\cite{techgap}. 
Closing this efficiency gap could improve science yield by a factor of several. For example, according to 
\citenum{stark2019exoearth}, the expected yields for characterized exo-Earths with coronagraphs baselined in the LUVOIR\cite{Luvoir19} and HabEx\cite{Gaudi20} final reports would be approximately 8 and 24 for a 6\,m inscribed-diameter, on-axis and off-axis aperture, respectively. If the coronagraph throughput, inner working angle (IWA), and tolerance to stellar angular size on these designs can be improved to theoretical limits, then these yields become, respectively, 44 and 48\cite{Belikov21}. 
This corresponds to a factor of 2 improvement in yield for off-axis apertures, and a factor of approximately 5 for on-axis apertures. 

Roughly speaking, closing the performance gap between coronagraphs for on-axis and off-axis apertures enables a factor of approximately 2-3 improvement in yield. Improvements in throughput, IWA, and tolerance to stellar angular size (and equivalently, to low-order telescope instabilities such as tip/tilt) achieve another factor of approximately 2 (although caution should be taken with interpreting the improvements due to IWA because the yield calculation does not take into account statistical challenges of small IWAs described in \citenum{Mawet14}). In addition, improvements in spectral bandwidth as well as non-coronagraphic elements (such as telescope mirror coatings and detector quantum efficiency) can lead to further yield improvements. 

In general, any performance improvement that is equivalent to increasing aperture diameter (such as improving photon flux at the detector, and IWA) is one of the most effective levers to improve HWO and reduce its risk. This is because yield depends more strongly on aperture diameter than almost any other mission parameter. 
Notably, yield is a very weak function of contrast, once contrasts approach 1e-10 levels\cite{stark2019exoearth} 
due to contrast being dominated by astrophysical backgrounds such as zodiacal and exozodiacal light. Because yield is such a strong function of aperture diameter, small improvements in yield are equivalent to large relaxations on aperture diameter requirements, should downscoping become necessary due to cost constraints. In addition, larger yields mitigate the risk of $\eta_{Earth}$ being smaller than is currently estimated. Another very effective lever to reduce risk is improving coronagraphic tolerance to telescope instabilities (without sacrificing IWA), because telescope instability is one of the most risky and expensive aspects of HWO.

In addition to reducing mission risk and improving expected yields, or quantity of exoplanets, it is important to also consider the quality of the science on those exoplanets. For the latter, it is critical to be able to do spectroscopy in a wide band and with high spectral resolution, as well as polarimetry. Polarimetry at small IWAs enables exciting science such as measuring Rayleigh scattering and ocean glints\cite{vaughan2023chasing}. However, this requires sacrificing some efficiency in practice. Improvements in all these aspects, without too much sacrifice, are important in order to maximize the quality of the science in addition to quantity.

All of these improvements are theoretically possible with further design and development of the coronagraph instrument on HWO. What is not clear yet is the most promising path to actualize those improvements, as well as the engineering effort required. One promising path, however, is photonic technologies. A key feature of photonic technologies is that with sufficiently advanced engineering, they have unlimited potential in the performance they can achieve (limited only by fundamental physical law). This is because there are photonic architectures that can implement a “universal” optical element\cite{miller2013self}. 
By contrast, traditional coronagraph designs that achieve physics limits do not yet exist, even assuming arbitrarily advanced engineering, at least without using a prohibitively large number of optical elements (though traditional coronagraphic designs continue to improve and may still come close to physics limits in the future). 

Although the theoretical potential of photonic chips is very encouraging, it needs to be tempered by practical considerations such as current maturity and development effort. Therefore, we propose two different architectures hybridizing photonic technologies with a traditional coronagraph with the goal of combining the best features of both and striking an optimal balance between performance and practical feasibility. 





\subsection{Background}

A fully integrated photonic system offers several crucial advantages over any current classical coronagraphic high-contrast imaging system due to its integrated architecture. This on-chip design is compact as well as thermally and mechanically stable, relaxing spacecraft stability requirements. A fully photonic solution reduces design complexity because it avoids potential additional limitations when combining multiple systems together.

The miniaturization in this fully photonic design allows for cost-efficient replication enabling the coverage of a large bandwidth with several channels. The theoretical limit intrinsic to photonic solutions is an IWA down to 0.5\,$\lambda/D$, which covers significantly more of the field of view (FOV) compared to current technologies. 

The integrated nature of this design allows a single astronomical instrument to hold multiple technologies, because of the shared hardware on one chip. The photonic architecture inherently separates by modes which can directly be used for different applications such as wavefront sensing and nulling.

The primary anticipated drawbacks of a fully photonic design are mainly due to the early developmental state of many of these technologies. Currently, one limitation of photonic capabilities is the coupling of the light into waveguides, and the light collection at the output, resulting in loss of light which could prove crucial for high-contrast imaging applications on space telescopes. Another challenge is leakage due to waveguide crossings\cite{astrophotonics_roadmap,cvetojevic20223beam,chingaipe}. Additionally, photonic nulling performance is greatly reduced with extended sources; a limitation which would significantly impact this proposed system's ability to detect exo-Earths\cite{benisty}. Furthermore, coupling into a single single-mode fiber (SMF) becomes less efficient for wide FOV coverage, pushing applications of photonic lanterns or other multi-mode fiber technologies\cite{jovanovic,Leon-Saval}. Another critical challenge is the transmission efficiency and therefore sensitivity of fiber interferometry\cite{Stoll}. Birefringence of waveguides is another challenge that needs to be mitigated\cite{gatkine}.  Overall, a fully photonic solution would require advancing current technology to reach $10^{-10}$ contrast capabilities for the entire FOV across a large bandwidth, which has not yet been demonstrated.


\section{Suggested Photonic-Coronagraphic Systems}

\begin{figure}[ht]
   \begin{center}
   \begin{tabular}{c}
   \includegraphics[width=14cm]{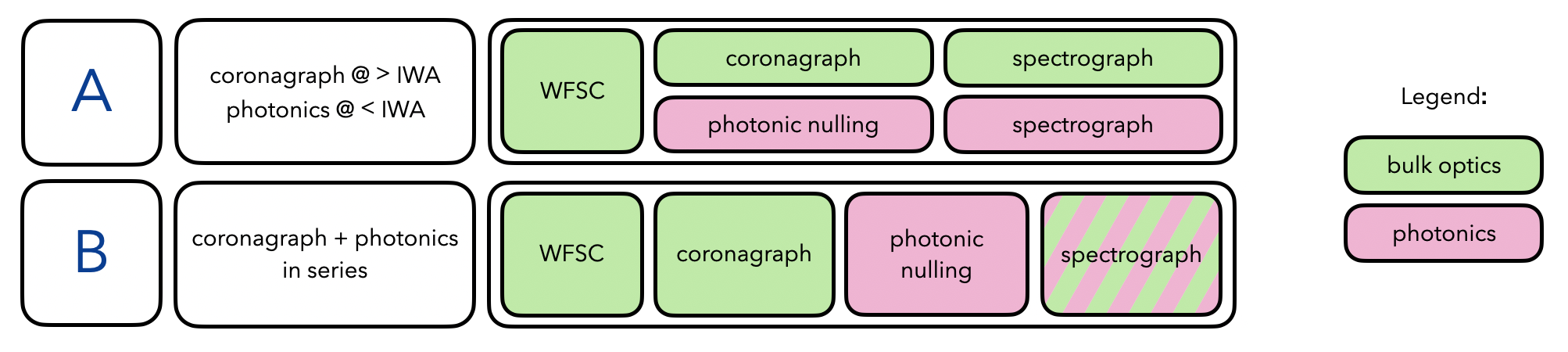}
   \end{tabular}
   \end{center}
   \caption[Fig] 
   { \label{fig:block} Two suggested photonic-coronagraphic hybrid systems.}
   \end{figure}

We propose two possible solutions to integrate photonics into existing coronagraphic systems. These hybrid solutions can leverage the advantages of new emerging photonics capabilities and address the limitations of existing coronagraph systems. Both solutions target the goal of significantly increasing potential exo-Earth yield by lowering the effective IWA and allowing detection of planets at close separations to their host star. A block diagram of both systems is shown in Figure~\ref{fig:block}.

\subsection{Solution A: Coronagraph for large separations, photonics within the IWA}

The first proposed system combines a classical coronagraph with photonic technologies in parallel. A classical coronagraph will image planets at larger angular separations. Any light rejected by the classical coronagraph however, is injected into a photonic coronagraph, allowing it to be used for photonic wavefront sensing or even a photonic coronagraph. This solution effectively separates the light at small and large angular separations with the classical coronagraph, allowing each coronagraph to work in their own optimal regime.

This solution has little impact on the operation of the classical coronagraph, thereby allowing for easier integration as an experimental module in an existing coronagraphic system. Initial implementations could focus on photonic wavefront sensing rather than a photonic coronagraph, which will likely be easier to implement with current technological limitations. However, if one wants to optimize the performance of the entire system as a coronagraph, then the IWA of the classical coronagraph becomes an important parameter. Larger IWAs now make the classical coronagraph more robust against low-order wavefront aberrations, while simultaneously allowing the photonic coronagraph to detect planets within this IWA.

\begin{figure}[t] 
   \begin{center}
   \begin{tabular}{c}
   \includegraphics[width=.9\textwidth]{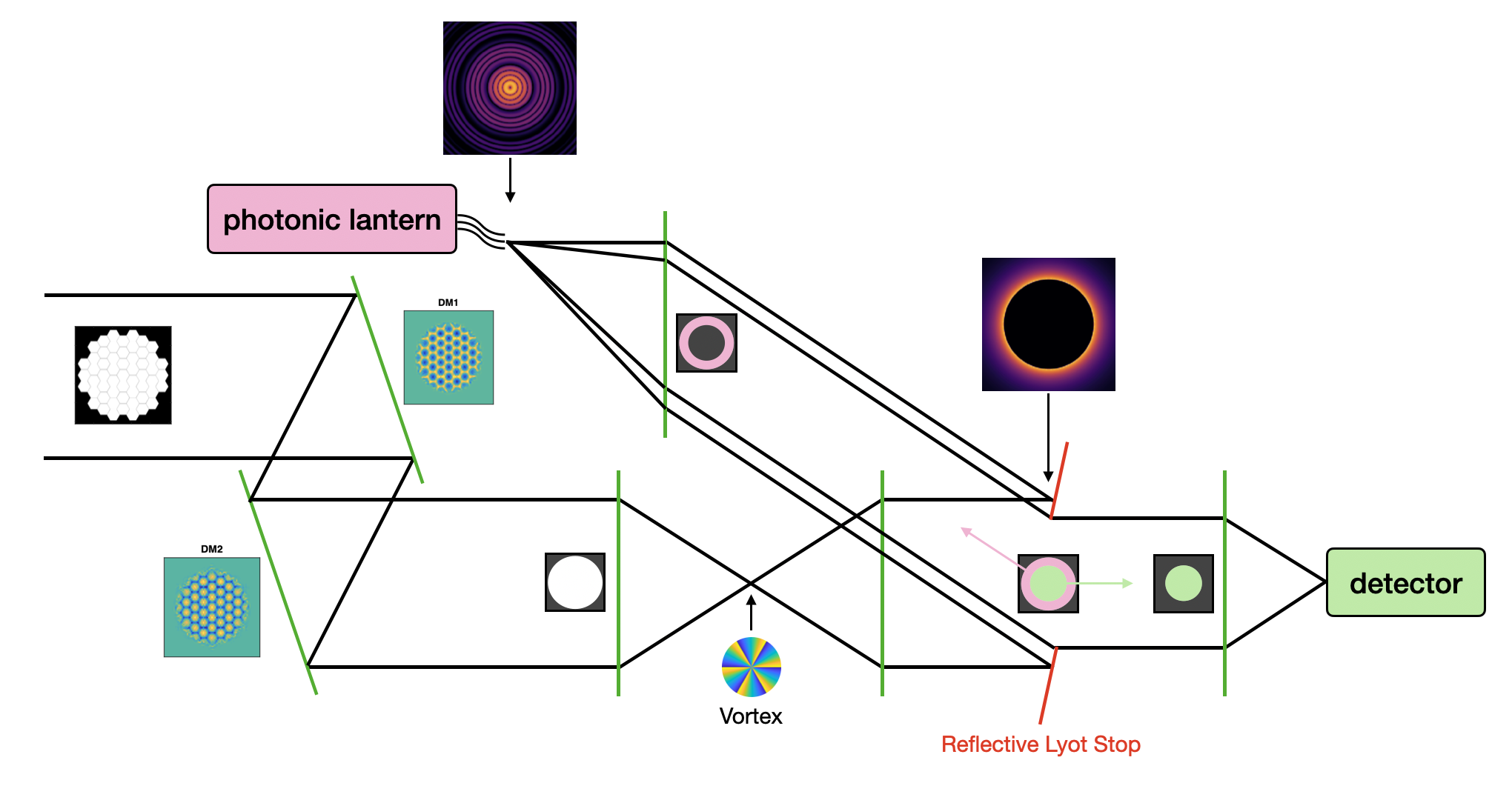}
   \end{tabular}
   \end{center}
   \caption[Fig] 
   {\label{fig:reflectivelyot} Schematic of system A using a reflective Lyot stop to recollect the light rejected by the coronagraph, and feeding it into a photonic lantern.}
   \end{figure}

Here, we simulate an example system for this solution. A schematic overview of this system is shown in Fig.~\ref{fig:reflectivelyot}. We use an apodized vortex coronagraph for the LUVOIR-B telescope pupil\cite{juanola2022modeling}. The light is apodized by two deformable mirrors (DMs) and a pupil mask, and then focused onto a vortex coronagraph. An idealized vortex phase mask is used. The light is then collimated onto the circular Lyot stop. The light inside the Lyot stop is transmitted and focused on the detector, which forms our classical coronagraph. The light outside the Lyot stop however contains all light inside the IWA of the classical coronagraph. We apply a vortex phase pattern with opposite charge to that of the focal-plane phase mask onto the Lyot stop. This cancels the vortex phase of the focal-plane mask. Then, the reflected light is focused onto a photonic lantern. We use a simple projection onto Laguerre-Gaussian
modes for this proof-of-concept simulation.

\begin{figure}[ht] 
   \begin{center}
   \begin{tabular}{c}
   \includegraphics[width=.4\textwidth]{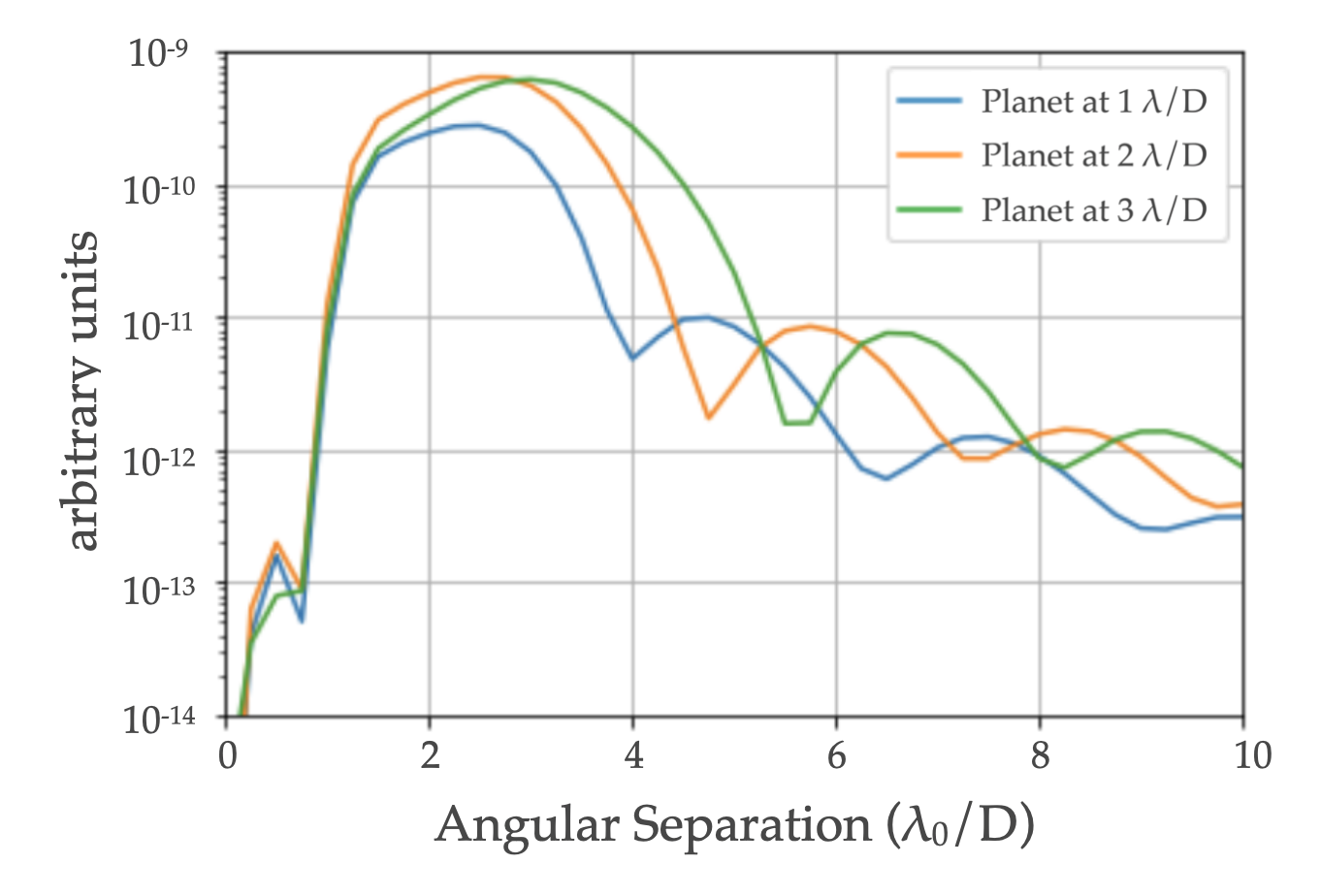}\includegraphics[width=.6\textwidth]{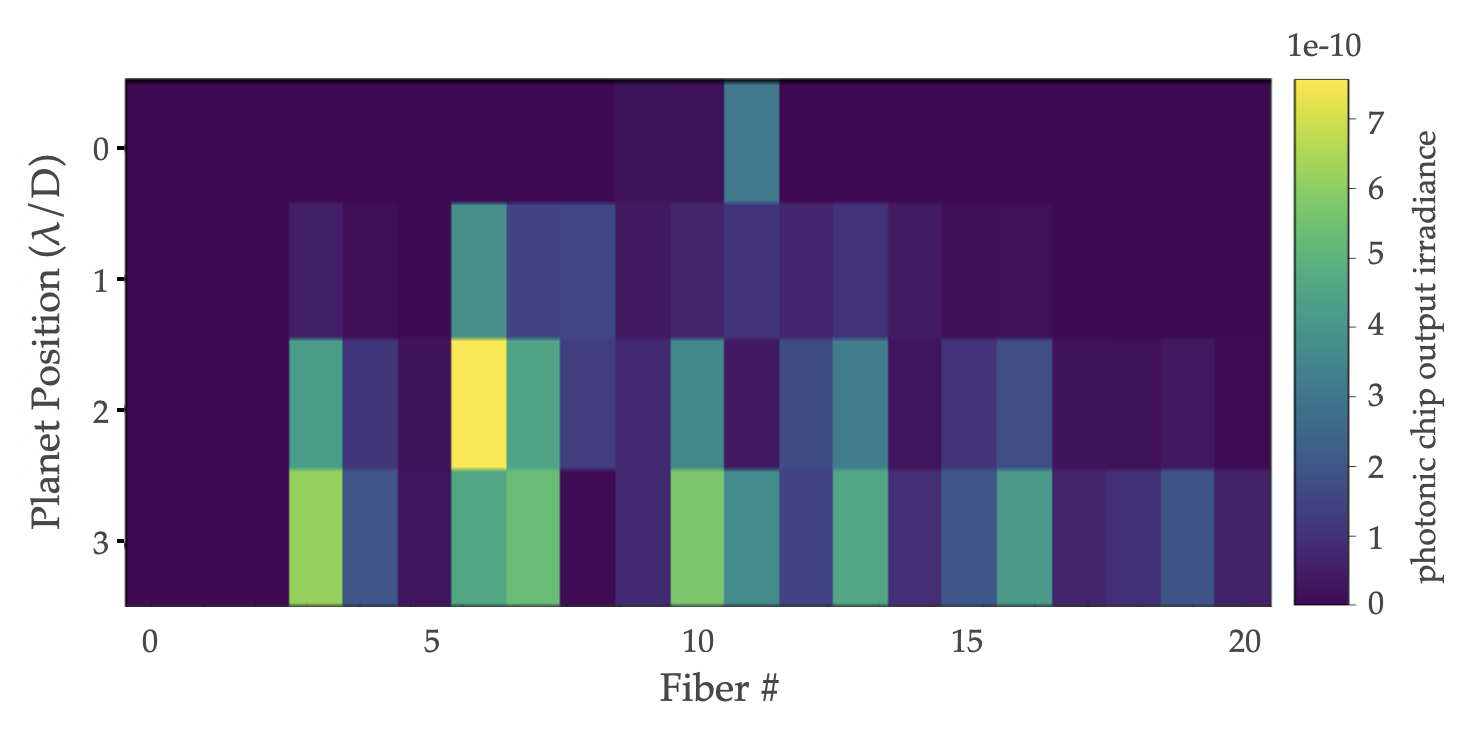}
   \end{tabular}
   \end{center}
   \caption[Fig] 
   { \label{fig:systema} Left: Simulated output of 21 fibers coupled with the light within the IWA, reflected off the Lyot stop. This plot shows the integrated fiber coupling efficiency with the input plane/star normalized in each case. Right: Projected modes for different positions of the source.}
   \end{figure}

Figure~\ref{fig:systema} shows the simulated output of 21 fibers coupled with the reflected light off the Lyot stop, corresponding originally to light from within 3\,$\lambda/D$ (IWA). The right plot shows the projected modes as a function of the position of the source. When on-axis, the vast majority of the light is concentrated in fibers with a radially-symmetric mode (i.e., fibers \#9, \#10 and \#11), while off-axis sources' light is distributed in other fibers. Therefore, the photonic lantern already provides some suppression of starlight, even without further processing.

Potential challenges of this hybrid system include maximizing the coupling efficiency into the photonic system. This requires matching and/or optimzing the F-number at the input of the photonic lantern. Furthermore, the requirements on mode-selectivity of the photonic lantern, manufacturing errors in the photonic lantern itself and scattering problems at high spectral resolutions inside the lantern are unknown as of this writing.

\subsection{Solution B: Coronagraph and photonics in series}

\begin{figure}[t] 
   \begin{center}
   \begin{tabular}{c}
   \includegraphics[width=.8\textwidth,trim={2.4cm 8cm 3.3cm 6.0cm},clip]{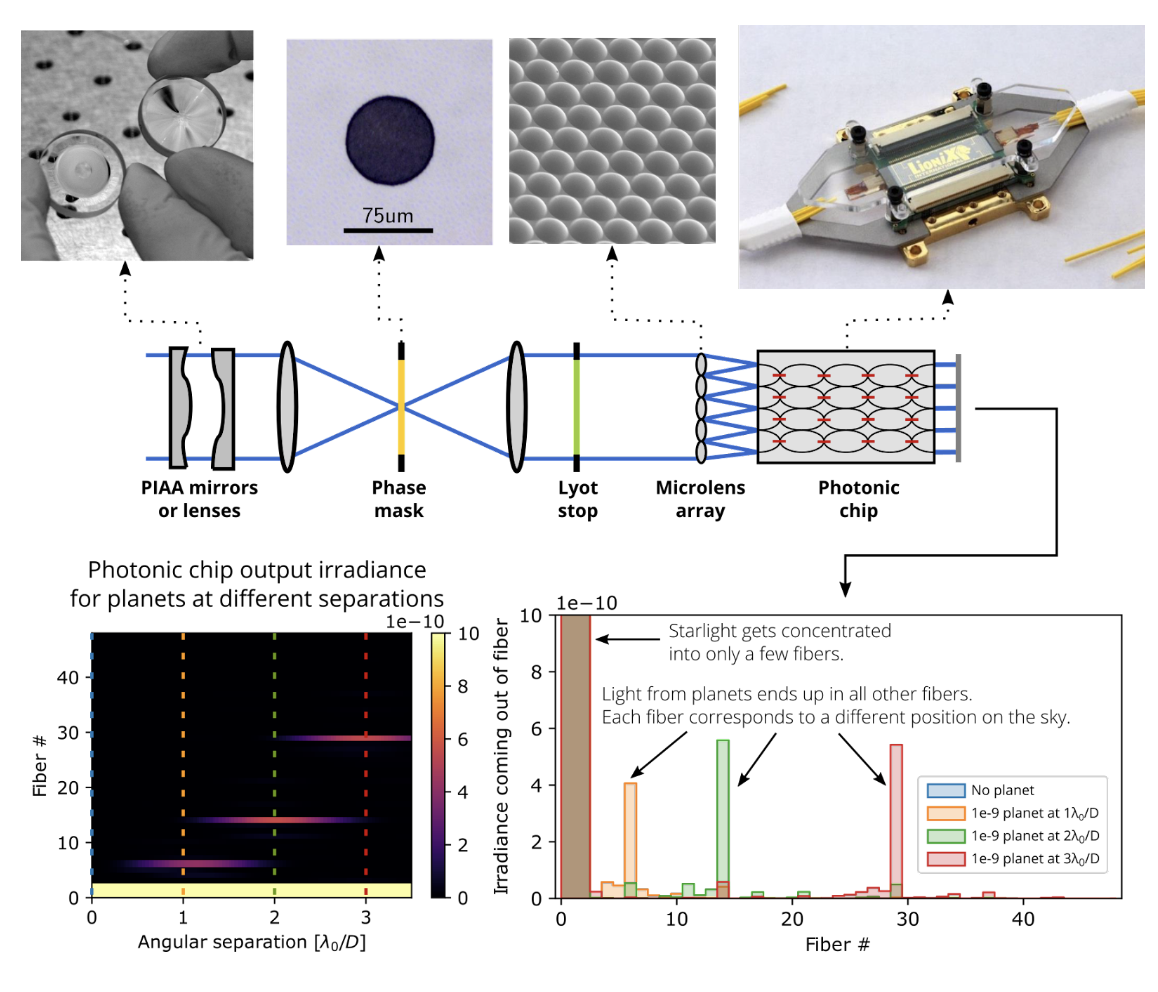}
   \end{tabular}
   \end{center}
   \caption[Fig] 
   {\label{fig:photonic_coronagraph_series} Schematic of system B combining a traditional coronagraph and photonics in series. The coronagraph shown is the PIAACMC combined with a photonic chip enabling photonic nulling.}
   \end{figure}

The second proposed system combines the coronagraph and photonics in series. By first implementing a traditional coronagraph with an extremely small IWA to suppress most of the on-axis starlight, at the cost of being extremely sensitive to wavefront aberrations and chromaticity, and then feeding the signal into a photonic chip at a downstream focal plane, we can potentially offload disadvantages of each technique onto each other, leading to a better design of the full system. Furthermore, since the final system is photonic, a spectrograph -- either a traditional or photonic one -- can be attached with little loss in throughput. A schematic overview of this solution is shown in Fig.~\ref{fig:photonic_coronagraph_series}.

Compared to solution A, solution B reduces the starlight suppression requirements on the photonic component. While for solution A, all the starlight is injected into the photonic component, here the starlight is already mostly suppressed by the upstream conventional coronagraph, leading to a reduction in scattering problems and mode-selectivity problems. Say for example that the upstream coronagraph can reach contrasts of $10^{-6}$, then the photonic coronagraph component only needs to filter this light by $10^{-4}$ itself to reach the required cumulative $10^{-10}$ requirement. A big downside of solution B is that all the light is injected into the photonic component, meaning that the outer working angle is limited by the number of fibers in the photonic lantern or integrated photonic chip, yielding typically 2-3\,$\lambda/D$ outer working angles for currently-foreseen technology (equivalent to 19--37 fibers/modes).

For our proof-of-concept simulation, we simulate a Phase-Induced Amplitude Apodization Complex Mask Coronagraph (PIAACMC) for a 15\% obscured circularly-symmetric aperture for a 20\% spectral bandwidth. We use a simple $0.55~\lambda_0/D$ radius phase dimple, which shifts the light by $\pi$ radians, as our focal-plane mask. A PIAACMC apodizer is then optimized for this system which suppresses the light for the center wavelength. After the Lyot stop, the light is focused by a microlens array, and then injected into a photonic chip. This photonic chip contains a mesh of Mach-Zehnder interferometers (MZIs), which can be tuned to provide any arbitrary matrix-vector multiplication to the incoming mode coefficients. We optimize this matrix to filter out most of the off-axis light due to chromaticity in as few fibers as possible (in this example, 4 fibers/modes), then make the rest of the fibers selective to positions on the sky. Finally, we observe the intensity in each of the output waveguides. No tolerancing was performed on each of the simulated components.

   \begin{figure}[ht] 
   \begin{center}
   \begin{tabular}{c}
   \includegraphics[width=.55\textwidth]{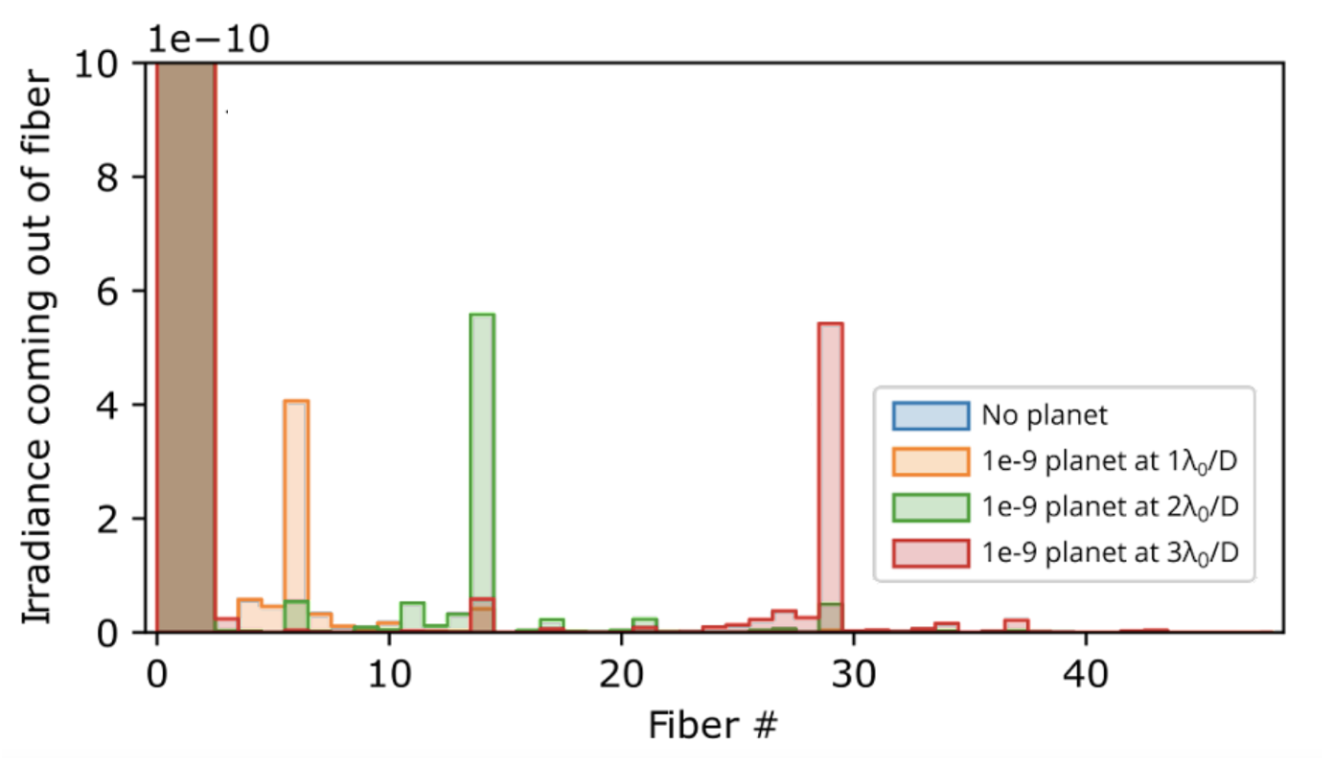}\includegraphics[width=.4\textwidth]{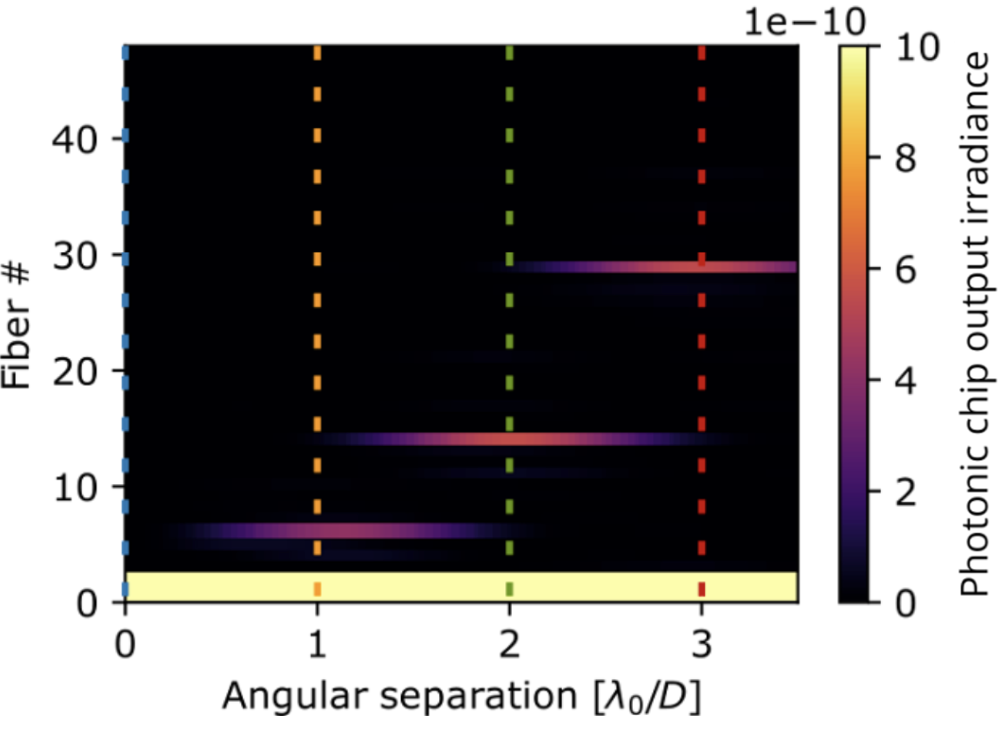}
   \end{tabular}
   \end{center}
   \caption[Fig] 
   { \label{fig:systemb} Left: Simulated output of 50 fibers coupled with the residual light from the traditional coronagraph. Right: projected modes for different positions of the source. The on-axis light is concentrated in a few fibers and can be removed without significantly decreasing the planet throughput, concentrated in different fibers. The coloured vertical lines correspond to the same colors on the left subplot.}
   \end{figure} 

The simulated output of the proof-of-concept simulations are shown in Fig.~\ref{fig:systemb}. We simulate a system with no planet, and three with a $10^{-9}$ planet at different angular separations. Clearly, most of the on-axis light is indeed filtered into the first 4 fibers, with the rest of the fibers being dedicated to planet light. Additionally, at 1~$\lambda_0/D$, most of the light is injected in fiber \#6, which is the fiber at that location on the sky. In the end, about 40\% of planet light is injected into this single fiber, equivalent or better than the core throughput of many traditional coronagraphs. The right subplot of Fig.~\ref{fig:systemb} shows the output of each fiber in a 20\% spectral bandwidth for our obscured telescope pupil as function of angular separation of our simulated $10^{-9}$ planet. The light from the planet gradually switches from one fiber to the other, depending on its angular separation.

One major drawback of solution B is that it needs to be more tightly integrated with the rest of the instrument. That is, the ability for the photonic chip to concentrate the leaked starlight of the conventional coronagraph into as few fibers as possible, likely hinges directly onto the design of the traditional coronagraph itself. Therefore, one likely cannot simply take an existing PIAACMC coronagraph and attach a photonic chip. This makes it more difficult and expensive to test this solution on a full instrument, lessening the chances of being integrated as a technology demonstration subsystem. 

Finally, as with solution A, there are many unknowns with solution B as well. One of the major challenges will be fine-tuning the system injecting light into the integrated photonic chip. Additionally, as the design space of this solution is unknown, its ultimate performance and/or capabilities with more complicated, non-circular telescope pupils are also unknown. Additionally, tolerancing of the MZIs on the photonic chip might create problems to achieve the angular selectivity needed for this design. Finally, inherently to the design, the outer working angle will be limited to that of the photonic components used, typically 2-3~$ \lambda/D$ for the currently foreseen 19--37 fibers/waveguides.





\section{Discussion}
\label{sec:discussion}

A fully photonic solution would offer many advantages over current classical coronagraphic high-contrast imaging systems. However, the level of technology maturation required to implement such a  solution for NASA's next flagship space telescope mission might be ambitious. Toward the effort of increasing exoplanet yield, and more specifically tackling improving the IWA, this study provides methods to leverage photonic capabilities in tandem with classical coronagraphs. The two proposed systems suggest either splitting the signal spatially or placing the two technologies in series. The proposed solution A allows for a traditional coronagraph to suppress starlight outside a large IWA and a photonic infrastructure to null starlight and isolate planet light within the IWA. The proposed solution B relaxes requirements on the coronagraph by sending all the signal output from the conventional coronagraph to the photonic chip for nulling and spectroscopy across the entire FOV.

Both systems present examples of ways to leverage advantages of current coronagraph advancements and new photonic development. 
The simulations for both methods showed potential to successfully couple planet signal into different fiber modes at the $10^{-10}$ level. However, implementing these hybrid systems presents several unique challenges for both future coronagraph development and photonics development still.

Although the area of photonics for astronomical applications is fairly new and has a lot of potential room for improvement, improvements to traditional coronagraphs would also enable improved performance and assist in achieving requirements for the hybrid designs proposed. One important component that can help improve the overall yield is an achromatic and polarization-independent focal-plane mask. Combined with a hybrid photonic solution it can alleviate the contrast requirements of the photonic chip by orders of magnitude. For instance, manufacturing better (more achromatic) focal-plane masks could theoretically decrease the number of modes required to be suppressed by the photonics from 10 to 3 or 4. 

On the photonics side, manufacturing constraints including fiber efficiency would play a large role in the performance of a hybrid system. The simulations in this study assume the photonic chips can be modeled with a perfectly invertible matrix and can map to output channels suppressing the starlight as desired. Realistically, however, manufacturing constraints might lead to optimization restrictions (as seen in traditional coronagraph design) which do not allow for this perfect assumption. As these technologies are further developed, these factors must be incorporated into simulated performance to gain more realistic yield calculations.

With targeted technology maturation and development both for current coronagraph technologies and upcoming photonic chips, high performing and reliable testbed environments will also be required to measure performance and push the limits of these components. Both photonic components, including integrated chips and photonic lanterns, as well as new increasingly complex coronagraphs will require testbed facilities to identify limitations and demonstrate performance.

Overall, combining new photonics capabilities with existing coronagraph technologies might reach smaller IWA, improve chromaticity, and relax the requirements on the coronagraph. As the fundamental limits of implementing coronagraph designs and photonic designs are better understood, hybrid systems like the ones presented here might offer unique advantages for future space telescopes.

\acknowledgments 

The 2023 Optimal Exoplanet Imagers workshop, which sparked the
work presented in this manuscript, was made possible thanks to the
logistical and financial support of the Lorentz Center, Leiden, The
Netherlands. The research presented in this paper was initiated at
a workshop held in Leiden and partially supported by NOVA (the
Netherlands Research School for Astronomy) and by the European
Research Council (ERC) under the European Union’s Horizon 2020
research and innovation programme (grant agreement 866001 -
EXACT). SRV acknowledges funding from the European
Research Council (ERC) under the European Union’s Horizon 2020
research and innovation program under grant agreement № 805445.
SLC acknowledges support from an STFC Ernest Rutherford Fellowship. TDG acknowledges funding from the Max Planck ETH
Center for Learning Systems. KB acknowledges support from NASA
Habitable Worlds grant 80NSSC20K152, and previous support for
related work from NASA Astrobiology Institute’s Virtual Planetary
Laboratory under Cooperative Agreement NNA13AA93A. IL
acknowledges the support by a postdoctoral grant issued by the Centre National d’Études Spatiales (CNES) in France. PB, IL, and YG
were supported by the Action Spécifique Haute Résolution Angulaire
(ASHRA) of CNRS/INSU co-funded by CNES. EHP is supported by
the NASA Hubble Fellowship grant \#HST-HF2-51467.001-A awarded
by the Space Telescope Science Institute, which is operated by the
Association of Universities for Research in Astronomy, Incorporated,
under NASA contract NAS5-26555. LA and EC acknowledge funding from the European Research Council (ERC) under the European
Union’s Horizon Europe research and innovation programme (ESCAPE, grant agreement 101044152). OA and LK acknowledge
funding from the European Research Council (ERC) under the European Union’s Horizon 2020 research and innovation programme
(grant agreement № 819155). SYH was funded by the generous support of the Heising-Simons Foundation. RJP is supported by NASA
under award number 80GSFC21M0002. OHS acknowledges funding
from the Direction Scientifique Générale de l’ONERA (ARE Alioth).
This research has made use of NASA’s Astrophysics Data System Bibliographic Services and the SIMBAD database, operated at CDS, Strasbourg, France.

\bibliography{bib}
\bibliographystyle{spiebib}

\end{document}